\documentclass[aps,prb,floatfix,preprint]{revtex4}
\usepackage{graphicx}
\usepackage{amsmath}
\usepackage{amssymb}
\usepackage{bm}
\usepackage{longtable}
\usepackage{array}
\usepackage{xspace}
\usepackage{multirow}
\usepackage{tensor}
\usepackage{algorithm,algpseudocode}
\usepackage{adjustbox}
\usepackage{mathtools}
\usepackage{mathrsfs} 
\usepackage{amsmath} 
\usepackage{float}     

\newcommand{\Eq}[1]{Eq. {\eqref{#1}}}
\newcommand{\icm}{\ensuremath{\mathrm{cm}^{-1}\;}}

\newcommand{\pot}[1]{\ensuremath{^{\left(#1\right)}}}

\newcommand{\kl}[1]{\ensuremath{\left(#1\right)}}

\newcommand\norm[1]{\left\lVert#1\right\rVert}

\begin{document}

\author{Sangeeth Das Kallullathil\renewcommand{\thefootnote}{\alph{footnote}}\footnote{Electronic address: 17sdk2@queensu.ca}}
\author{Tucker Carrington Jr.\renewcommand{\thefootnote}{\alph{footnote}}\footnote{Electronic address: Tucker.Carrington@queensu.ca}}
\affiliation{Chemistry Department, Queen's University, Kingston, Ontario K7L 3N6, Canada}
\date{\today}

\title{Computing vibrational energy levels by solving linear equations using a tensor method      with an imposed rank}

\begin{abstract}

  Present day computers do not have enough memory to store the high-dimensional tensors required when using 
  a direct product basis to compute vibrational energy levels of a polyatomic molecule with more than about 5 atoms.   
    One way to deal with this problem is to represent tensors using a tensor format.  In this paper, we use CP format. 
Energy levels are computed by building a basis from vectors obtained by solving linear equations.     The method can be thought of as a CP realization of a block inverse iteration method with multiple shifts.   The CP rank  
of the tensors is fixed and the  linear equations are solved with an Alternating Least Squares method.   There is no need for rank reduction,  no need for orthogonalization, and tensors with rank larger than the fixed rank used to solve the linear equations are never generated.   The ideas are tested 
by computing vibrational energy levels  of a 64-D bilinearly coupled model Hamiltonian and  of acetonitrile(12-D).

\end{abstract}
\maketitle

\section{Introduction}
\label{intro}

The most general and systematic strategy for computing a vibrational spectrum from a potential energy surface (PES), without making approximations, is to expand wavefunctions in a basis and solve a matrix eigenvalue problem obtained from a variational method.\cite{Carney78,Te,CaHa,Ca17}   
Implementing this strategy is computationally costly because       using established methods both the computer time and the computer memory required increase rapidly  with the size of the basis.
A direct product basis, each of whose functions is a product of univariate functions, 
\begin{align}
	 \Phi_{i_1,\cdots, i_D}(q_{1},...,q_D) =
	& 
	\varphi_{i_{1}}\pot{1} \kl{q_1}  
	\varphi_{i_{2}}\pot{2} \kl{q_2}
	\ldots  
	\varphi_{i_{D}}\pot{D} \kl{q_D}  
	\label{sopb1}   
\end{align}  
has the important advantage of simplicity.   
In this paper, $ D $ is the number of coordinates and the  coordinates are $ q_c $, $ c= 1, \cdots, D $. The obvious disadvantage of a direct product basis is its size, $ n^D $.  
$ n $ is a representative value of $ n_c $, $ c=1, \cdots, D $, which is the number of 1-D basis functions for coordinate $ q_c $.
To simplify the notation we use $ n $ throughout the paper, but it is simple to use different $ n_c $ for different coordinates.
Despite its size, it is possible to use a direct product basis by exploiting its structure to facilitate the evaluation of matrix-vector products (MVPs) and employing iterative methods to solve the matrix eigenvalue problem.
\cite{brca,Yu02,LiCa00,CFSMFC12} 
  For molecules with more than five atoms, 
it becomes impossible to store (even) vectors with $ n^D $ components and one needs new ideas  to compute spectra.

In this paper, we introduce a new tensor method for using iterative methods  with a direct product basis.  %
 There is no need to store  vectors with $ n^D $ components.   To develop methods whose memory cost does not scale exponentially we have used what mathematicians call canonical polyadic (CP) format.\cite{CPkolda,CPHitch,CPB,CPpnas}   
The first such method was the Reduced Rank Block Power Method (RRBPM).  
\cite{CPLec}
It uses sum-of-product (SOP) basis functions, $ 	F_{}(q_{1},...,q_D) $,    that are linear combinations of the functions of a direct product basis,    
\begin{align}
	F_{}(q_{1},...,q_D)   %
	& = \sum_{i_1=1}^{n} \ldots \sum_{i_D=1}^{n}
	F_{i_1, \ldots, i_D} \ 
	\varphi_{i_{1}}\pot{1} \kl{q_1}  
	\varphi_{i_{2}}\pot{2} \kl{q_2}
	\ldots  
	\varphi_{i_{D}}\pot{D} \kl{q_D}  
	\label{sopb}   
\end{align}  
Each basis function is a sum of $ R $ products and  it is not necessary to store $ F_{i_1, \ldots, i_D} $ as $ n^D $ numbers.   Instead,  one exploits
\begin{align}
\mathbf{F}  = 
\sum_{\ell=1}^{R} 
\bigotimes_{c=1}^D    \mathbf{f}_{c,\ell}
\label{ftensor}
\end{align}
and stores $ RD $ vectors $ \mathbf{f}_{c,\ell}  $.  $ R $ is the number of terms in \Eq{ftensor} and is called the rank. 
$ \mathbf{f}_{c,\ell}  $ has  only $ n $ components.   The memory cost scales linearly with $ D $.   In terms of components \Eq{ftensor} is 
\begin{align}
{F}_{i_1,i_2,..,i_d} 
& = 
\sum_{\ell=1}^{R} 
\prod_{c=1}^{D} f_{i_c}^{(c,\ell)}    %
\end{align}
The Multiconfiguration Time-dependent Hartree (MCTDH) method can also be considered a tensor method.\cite{Meyer90,Manthe92,potfit}  
 It, however, uses not CP format but Tucker format.\cite{CPkolda} 
   Because it uses  Tucker format, the memory cost of an MCTDH calculation scales exponentially with $ D. $
The advantages of using CP format for electronic structure problems have also been recognized. \cite{Jerke18}   
Multilayer-MCTDH  uses  basis functions that are products of multivariate
functions.\cite{WT3,M6,VeMe}  
It reduces the cost of an MCTDH calculation and is widely used when the PES is a SOP.    
 The 
density matrix renormalization group (DMRG) method is often used to calculate a ground state energy (but not the corresponding wavefunction). \cite{US11} 
Related ideas  have also been used to compute a small number of vibrational\cite{BaRe17,BaRe19}  or electronic\cite{WoNe14} states.   
Larsson has used  a  DMRG-type optimization  algorithm in conjunction with multilayer trees.  \cite{La19}    

There also exist non-tensor iterative methods that make variational calculations possible for molecules with more than five atoms.   They all use a basis that is not a direct product.   One class of such methods begins, at least conceptually, with a direct product basis and prunes away unnecessary basis functions.   Some of these approaches work only if the PES has a special (e.g. SOP)
form\cite{Halo83,WaCa01,Po03,LaTa16},  
and some work with a general PES\cite{AvCa12,AvCa09,AvCa15}.   
Another class of such methods uses basis functions that are  products of multivariate  functions which  are eigenfunctions of sub-problems.\cite{brcacontr,WaCa02,WaCa18,LeLi04,FeBa20}     %
For methods of both classes what is tricky is devising a scheme for evaluating MVPs.   
Selected configuration interaction methods have also been used successfully for large molecules.\cite{BhBr21,FeBe21,GoPo05,CaLi03}

Tensor algorithms are only useful if the Hamiltonian operator has a compatible form.   To use the RRBPM, the Hamiltonian must be a SOP.  To use most MCTDH programs, the Hamiltonian must also be a SOP (however, there are non-SOP MCTDH calculations using Manthe's Correlation Discrete Variable Representation\cite{M3} or collocation\cite{WoCaIII}).   If the Hamiltonian does not have the required form it can, at the cost of some extra calculations, be replaced by an operator that is approximately equal to the original Hamiltonian, but does have the required form.\cite{JaeMe,Schr,MaCa,BrPr16,KoZh}
For many molecules with more than five atoms, the best available PES is a SOP. \cite{GaFort20}

Although the original RRBPM eliminates the memory problem  one confronts when doing a variational calculation, it requires a lot of computer time.  It is inefficient because the   (shifted)  power method requires many MVPs and because after each MVP it is necessary to reduce the rank of the output vector.   The rank of the output vector is $ RT $, the product of the rank of the input vector (R) and the number of terms in the SOP Hamiltonian, in this paper denoted by $ T. $
  Several related, but more efficient, ideas have been introduced.   One divides the full problem into a sequence of sub-problems each with a subset of the full $ D $ coordinates by imposing a tree structure.  
\cite{CPTree} 
This Hierarchical H-RRBPM was later improved by 
 intertwining the matrix-vector products and the rank reduction algorithm.  \cite{CPint}
Rakhuba et al. replaced the CP format used in the original RRBPM with the Tensor Train (TT) or Matrix Product State format and replaced the shifted power method eigensolver with a combination of  LOBPCG (locally optimal block  preconditioned conjugate gradient)    and inverse iteration.  \cite{TTVIB}
   Because PESs for molecules with more than five atoms are often determined in SOP form, they  convert a  CP PES tensor to TT.
 In Larsson's Tree  Tensor Network States method  a tensor tree is used and the Schr\"odinger equation is solved with a DMRG-type optimization algorithm.  In practice, he computes states  one by one.  \cite{La19}

 Like the method of Rakhuba and Oseledets   \cite{TTVIB} our method is designed to simultaneously calculate all states in a given energy window.   
 Compared to the RRBPM and the method of  Rakhuba et al.,
  the method of this paper has important advantages:   no orthogonalization is necessary; no rank reduction (``rounding") 
  is required.  Rank reduction is  
  the most time consuming step in most  RRBPM calculations. 
  In the RRBPM, rank reduction is required after each MVP, after each orthogonalization, and after each update.\cite{CPLec}  
   The cost of the new method is much lower than the cost of the RRBPM.  
  The method has    two components:  1) eigenvalues and eigenvectors of the direct  product representation of the Hamiltonian operator, which is here  denoted $ \mathbf{H} $, are determined by projecting into a basis of filtered vectors; 2)  the filtered vectors are computed by exploiting the SOP character of the Hamiltonian operator and the direct product nature of the primitive basis.   Component 1) somewhat resembles Block Inverse Iteration,\cite{chatelin}
    the Filter Diagonalization method,\cite{wallN95,MTdirdel97,Man98,fdzslc}        
     and the Rational Krylov method\cite{Ruhe84}.     
  The  filtered vectors are  designed so that each is a linear combination of a small number of  eigenvectors of $ \mathbf{H} $  whose corresponding eigenvalues are
  in some selected energy range.        
  We call the method CP-MSBII.  It is a CP implementation of a Multiple Shift Block Inverse Iteration Method.       
Rank reduction is not required because filtered vectors are computed by solving linear equations while constraining the rank of the solution.  
 CP-MSBII   is tested by computing energy levels of 
a 64-D bilinearly coupled model Hamiltonian and acetonitrile.

\section{ Constraining the CP rank of the solution of a system of linear equations      }  

\label{cplin}

Ideal filtered basis  vectors  would be   determined (see section \ref{msbiie})   by solving,  
\begin{align}
	(\mathbf{H}-\sigma \mathbf{I}) \mathbf{F}^{ex} =\mathbf{s}~.  
		\label{lin}
\end{align}
Both $\mathbf{s} $ and $ \mathbf{F}^{ex} $ are tensors with $ n^D $ components. 
 In the CP-MSBII method (see section \ref{msbiie}), $  \mathbf{s}  $ is in CP format and has rank $ R $. 
  $ \mathbf{H} $ is an $ n^D \times n^D$ matrix representation of a SOP operator.
Rather than computing $ \mathbf{F}^{ex} $, we find  approximate solutions, henceforth denoted  $ \mathbf{F}$,
 by 
replacing $ \mathbf{F}^{ex} $  in \Eq{lin}  with $ \mathbf{F} $ in \Eq{ftensor}
 which is in    CP format, and requiring that  $ \mathbf{F} $    have rank $ R $.
 To determine  basis vectors from which one can compute accurate eigenvalues (see section    \ref{msbiie}) it is sufficient to solve \Eq{lin} approximately.   
    In this section, we explain how to compute   $ \mathbf{F} $       by minimizing 
  $||(\mathbf{H}-\sigma \mathbf{I})\mathbf{F}- \mathbf{s}    ||$.  
  An Alternating Least Squares (ALS)
  algorithm is used.    
  The ALS solution  is more accurate if the exact solution of \Eq{lin} is low rank.  This will be the case if $  \mathbf{s}  $ is a linear combination of a small number of the eigenvectors of   $ \mathbf{H} $ with the smallest eigenvalues.   In section \ref{msbiie}  
  we discuss choosing $  \mathbf{s}  $  to satisfy this requirement.

  Using ALS, we  sequentially determine   $\pmb{\mathcal{F}}_j$ for $ j=1,2, \cdots, D $, keeping       $\pmb{\mathcal{F}}_{c \ne j} $ fixed.   
$\pmb{\mathcal{F}}_c$ is defined  so that its  $\ell$th   column,   $\ell=1,...,R$,  
is  
$\mathbf{f}_{c,\ell}$  (see \Eq{ftensor}).    The number of elements in the   $\ell$th   column is   $ n$.  
$\pmb{\mathcal{F}}_c$ is a collection of parameters for coordinate $q_c$. 
The columns of $\pmb{\mathcal{F}}_c$  are not normalized.  
Computing   $\pmb{\mathcal{F}}_j$ for $ j=1,2, \cdots, D $  constitutes one sweep.  The most accurate possible ALS solution is obtained by doing many sweeps, however, we find that solutions determined with a single sweep are accurate enough.

The ALS algorithm we use is modelled on the one in Ref. 
\onlinecite{Alscp}.  
For any coordinate $ j $,  we define  
\begin{align} 
 \mathbf{W}^F_j(\pmb{\mathcal{F}}_{1},...,\pmb{\mathcal{F}}_{j-1},\pmb{\mathcal{F}}_{j+1},...,\pmb{\mathcal{F}}_{D}) \in \mathbb{R}^{n^D \times nR} ~, 
  \end{align}
 which  depends on the parameters of all coordinates except  $\pmb{\mathcal{F}}_{j}$, so that 
\begin{align}   \mathbf{F} = 
 \mathbf{W}^F_j vec(\pmb{\mathcal{F}}_{j})~.
 \label{WPeV}
\end{align}
$  vec(\pmb{\mathcal{F}}_{j}) $ is a $ nr  \times 1 $ column made by stacking the columns of 
$  \pmb{\mathcal{F}}_{j} $.
It is straightforward to show that     
\begin{align}
\begin{split}
\mathbf{W}^F_j(\pmb{\mathcal{F}}_{1},...,\pmb{\mathcal{F}}_{j-1},\pmb{\mathcal{F}}_{j+1},...,\pmb{\mathcal{F}}_{D})= (
{\mathbf{f}}_{1,1} \otimes  \cdots  \otimes {\mathbf{f}}_{j-1, 1}  \otimes  %
\mathbf{I}_j  \otimes {\mathbf{f}}_{j+1, 1} \otimes   \cdots  \otimes {\mathbf{f}}_{D, 1}
\big\vert
\\ 
{\mathbf{f}}_{1,2} \otimes   \cdots  \otimes {\mathbf{f}}_{j-1, 2}  \otimes
\mathbf{I}_j  \otimes {\mathbf{f}}_{j+1, 2} \otimes   \cdots  \otimes {\mathbf{f}}_{D, 2}
\big\vert
\\ 
\cdots
{\mathbf{f}}_{1,R} \otimes   \cdots  \otimes {\mathbf{f}}_{j-1, R}  \otimes
\mathbf{I}_j  \otimes {\mathbf{f}}_{j+1, R} \otimes  \cdots  \otimes {\mathbf{f}}_{D, R}   )
\end{split}  ~,
\label{w}  
\end{align}
where $  \mathbf{I}_j  $  is the $ n \times n $ identity matrix.     The vertical lines in \Eq{w} divide slices of the matrix each of which is $ n^D \times n $ matrix.   
$ \hat{H} $ is a SOP operator 
\begin{align}
	H(q_1,...,q_D)   
	& = 
	\sum_{t=1}^{T} 
	\prod_{c=1}^{D} \hat{h}_{c}^t{(q_c)}  ~.
	\label{hamilto}
\end{align}
In \Eq{hamilto},    $h_{c}^t{(q_c)}$ is a one-dimensional operator.   
 $(\mathbf{H}-\sigma \mathbf{I})$ is a CP matrix,  
\begin{align}
(\mathbf{H}-\sigma \mathbf{I})
& = 
\sum_{t=1}^{T+1} 
\bigotimes_{c=1}^D \mathbf{h}_{c}^t{(q_c)}. 
\label{hmat}
\end{align}
The $(T+1)$th term is $ \sigma \mathbf{I} $.
The right side of \Eq{lin} is written
\begin{align}
	\mathbf{s} = 
	\sum_{\ell=1}^{R} 
	\bigotimes_{c=1}^D    \mathbf{g}_{c,\ell}~.
	\label{gensor}
\end{align}
To determine $  \pmb{\mathcal{F}}_{j}$ we replace $ 
\mathbf{F}^{ex} $  in   
\Eq{lin} with $ \mathbf{F} $
and then use 
 \Eq{WPeV}  and multiply on the left by 
 $(\mathbf{W}^F_j)^T$  
\begin{align}
(\mathbf{W}^F_j)^T (\mathbf{H}-\sigma \mathbf{I}) \mathbf{W}^F_j vec(\pmb{\mathcal{F}}_{j}) = (\mathbf{W}^F_j)^T \mathbf{s}   ~.
\end{align}
Using \Eq{w}, ~ \Eq{hmat},    ~
\Eq{gensor},  and   $    \mathbf{s} = 
\mathbf{W}^G_j ~vec(\pmb{\mathcal{G}}_{j})~     $ 
 one finds a small linear system that can be solved for $ vec(\pmb{\mathcal{F}}_{j} ) $,
\begin{align}
\left[ \sum_{t=1}^{T+1} \left( \bigodot_{c\neq j}^D ({\pmb{\mathcal{F}}}_{c}^T \mathbf{h}_c^t 
{\pmb{\mathcal{F}}}_{c}) \right) \otimes \mathbf{h}_{j}^{t} \right ]   vec(\pmb{\mathcal{F}}_{j}) 
&= 
\sum_{\ell^{\prime\prime}=1}^{R} \left(   \bigodot_{c\neq j}^D   ({\pmb{\mathcal{F}}}_{c}^T {\mathbf{g}}_{c,\ell^{\prime\prime}})   \right)  \otimes  {\mathbf{g}}_{j,\ell^{\prime\prime}}   ~,   
\label{maineq}
\end{align}
where $\bigodot$ denotes  element wise (or Hadamard) multiplication.

When solving \Eq{maineq}, $ \pmb{\mathcal{F}}_{c \ne j} $ are fixed.   \Eq{maineq} is solved for $ j=1,2, \cdots, D $.  
  It has the form  $\mathbf{Ax}=\mathbf{b}$ and
$ \mathbf{A} $ is an $ nR \times nR $ matrix.  ALS converts a linear system with a   $ n^D  \times n^D $   matrix into a set of linear problems with 
 $ nR \times nR $ matrices.
To obtain the best possible solution, one would solve  \Eq{maineq},  for $ j=1,2, \cdots, D $  $,  N_{ALS} $ times, each time updating the 
 $ \pmb{\mathcal{F}}_{c \ne j} $ that  are fixed.   Instead we solve only once for each $ j $, i.e. $ N_{ALS} =1 $ (see Section IV).      %
If  \Eq{maineq} is written  $\mathbf{Ax}= \mathbf{b}$,  then elements of $  \mathbf{A} $ and   $  \mathbf{b} $ are  
\begin{align}
{A}(\ell^{\prime},i_j^{\prime},\ell, i_j)=\sum_{t=1}^{T+1} \left( \prod_{c\neq j}^D ({\mathbf{f}}_{c,\ell^{\prime}}^T \mathbf{h}_c^t {\mathbf{f}}_{c,\ell}) \right)  h_{i_j^{\prime},i_{j}}^{t},
\end{align}
and   
\begin{align}   
{b}(\ell^{\prime},i_j^{\prime})=\sum_{\ell^{\prime\prime}=1}^{R} \left[  \  \prod_{c\neq j}^D ({\mathbf{f}}_{c,\ell^{\prime}}^T  {\mathbf{g}}_{c,\ell^{\prime\prime}}) \right ] {g}_{i_{j}^{\prime}}^{(j,\ell^{\prime\prime})}.    
\end{align}
To avoid calculating $\mathbf{A}$ and $  \mathbf{b} $ from scratch for each $ j$, we  re-write these equations
\begin{align}
	{A}(\ell^{\prime},i_j^{\prime},\ell,i_j)=\sum_{t=1}^{T+1}M^{t}_{\neq j}(\ell^{\prime},\ell)  h_{i_j^{\prime},i_{j}}^{t} ,
\end{align} 
\begin{align}  
	{b}(\ell^{\prime},i_j^{\prime})=\sum_{\ell^{\prime\prime}=1}^{R}   P_{\neq j}(\ell^{\prime},\ell^{\prime\prime}) \  {g}_{i_{j}^{\prime}}^{(j,\ell^{\prime\prime})}~,
\end{align}
where 
\begin{align}
M^{t}_{\neq j}(\ell^{\prime},\ell)
&=
\prod_{c\neq j}^D ({\mathbf{f}}_{c,\ell^{\prime}}^T \mathbf{h}_c^t {\mathbf{f}}_{c,\ell}),
\end{align}
and 
\begin{align}
P_{\neq j}(\ell^{\prime},\ell^{\prime\prime})
&=
\prod_{c\neq j}^D ({\mathbf{f}}_{c,\ell^{\prime}}^T  {\mathbf{g}}_{c,\ell^{\prime\prime}})~.
\end{align}
Because it is small, we solve the linear system in \Eq{maineq} with methods of direct linear algebra.  The cost scales as $ (nR)^3 $.   It would also be possible to use a (preconditioned)  iterative linear solver.  When  $ T $ is small iterative solvers might be less costly.

A pseudocode for the implementation of the method is given in algorithm \ref{MSBII}.     
 $\mathbf{M}^t_{\neq j}$  and $\mathbf{P}_{\neq j}$ do not need to be calculated from scratch for each $ j. $
 They depend on $\pmb{\mathcal{F}}_{c \ne j}$.   
 After computing  $\pmb{\mathcal{F}}_{ j}$, it is used to update $ 
 M^{t}_{\forall c}(\ell^{\prime},\ell)$
 and  $
P_{\forall c}(\ell^{\prime},\ell^{\prime\prime})$.  
  The  cost of  constructing  $\mathbf{M}^{t}_{j}$    
   for one $ j $ is $n^2 R + n R^2$ 
   and  the  cost of building $\mathbf{M}^{t}_{\forall j}$ is $D(n^2R+nR^2)$.    
  Computing $\mathbf{A}$ therefore requires $(T+1) [D(n^2R+nR^2)+(nR)^2]$ operations.        %
$nR^2$ is the cost of making one $\mathbf{P}_{j}$ matrix and $D(nR^2)$ is the cost of making a $\mathbf{P}_{\forall j}$ matrix.  
   Computing $\mathbf{    
   {b}}$
   therefore costs 
      $D(nR^2)$  + $ R^2 n$.     
The algorithm requires the   matrices,
\begin{align}
M^{t}_{\forall c}(\ell^{\prime},\ell)
&=
\prod_{c=1}^D ({\mathbf{f}}_{c,\ell^{\prime}}^T \mathbf{h}_c^t {\mathbf{f}}_{c,\ell}),
\end{align}

\begin{align}
P_{\forall c}(\ell^{\prime},\ell^{\prime\prime})
&=
\prod_{c=1}^D ({\mathbf{f}}_{c,\ell^{\prime}}^T  {\mathbf{g}}_{c,\ell^{\prime\prime}}),
\end{align}
\begin{align}
M^{t}_{j}(\ell^{\prime},\ell)
&=
 {\mathbf{f}}_{j,\ell^{\prime}}^T \mathbf{h}_j^t {\mathbf{f}}_{j,\ell} ~,
\end{align}
and
\begin{align}
P_j(\ell^{\prime},\ell^{\prime\prime})
&=
{\mathbf{f}}_{j,\ell^{\prime}}^T  {\mathbf{g}}_{j,\ell^{\prime\prime}}.
\end{align}
 $\mathbf{M}^t_{\neq j}$ and $\mathbf{P}_{\neq j}$  are  calculated from  
\begin{align}
M^{t}_{\neq j}(\ell^{\prime},\ell)=M^{t}_{\forall c}(\ell^{\prime},\ell)/M^{t}_{j}(\ell^{\prime},\ell)
\end{align}
and 
\begin{align}
P_{\neq j}(\ell^{\prime},\ell^{\prime\prime})=P_{\forall c}(\ell^{\prime},\ell^{\prime\prime})/P_j(\ell^{\prime},\ell^{\prime\prime}).
\end{align}

	\section{Multiple shift block inverse iteration eigensolver}  
	
\label{msbiie}

	Our   filtered vectors are  computed by using ALS to solve successive systems  of linear equations.   A filtered vector, $  \tilde{\mathbf{	F}}^b_v  $, is designed to
	approximate  $ ( \mathbf{H} - \sigma^b_v   \mathbf{I}) ^{-P}  \mathbf{  s}^b_v. $ ~ For example, when $ P=2 $,

\begin{align}  
	\tilde{\mathbf{	F}}^b_v =    ( \mathbf{H} - \sigma^b_v   \mathbf{I}) ^{-1}     {	\mathbf{	L}}^b_v  ,
	\label{filtery}
\end{align}
where 
\begin{align}  
	\mathbf{  L}^b_v = \frac{\tilde{\mathbf{	L}}^b_v} {\norm{ \tilde{\mathbf{	L}}^b_v}},
\end{align}
and
\begin{align}  
\tilde{	\mathbf{	L}}^b_v =       ( \mathbf{H} - \sigma^b_v   \mathbf{I}) ^{-1}  \mathbf{  s}^b_v.
	\label{inter}
\end{align}
Both  $    	\tilde{\mathbf{	F}}^b_v $
and $     \tilde{	\mathbf{	L}}^b_v      $ are computed using the CP procedure of section \ref{cplin}.  
	The vectors are labelled by   $ b=1, \cdots, B $ and by     $ v=1, \cdots, S_b $.   
	 $ b $ labels a block of vectors.   %
	 $ S_b $ is the number of vectors in a block.  
	 The $ \sigma_v^b $ values are referred to as shifts.  $\mathbf{ s}^b_v $ is a start vector. 
	   The start vector is in CP format and has rank $ R. $      In some cases  $\mathbf{  s}^b_v   =     \mathbf{  s}^b_{v'}.  $ 	
	 We normalize $  \tilde{\mathbf{	F}}^b_v  $ to obtain $  {\mathbf{	F}}^b_v  $ and  build a basis from the  $  {\mathbf{	F}}^b_v  $.
 The basis vectors are not orthogonal.
Once  the filtered vectors are known,  we project into the space the  
filtered vectors span, 
 and solve a generalized eigenvalue problem,
\begin{align}
	\mathbf{H}^{{\mathscr{F}}}\mathbf{U}=\mathbf{S}^\mathscr{F}\mathbf{U}\mathbf{E}~,
	\label{geneig}
\end{align}
to determine eigenvalues and eigenvectors of 
$ \mathbf{H}  $. 
  In \Eq{geneig},
 $\mathbf{H}^{{\mathscr{F}}} = {\pmb{\mathscr{F}}}^{T} \mathbf{H} \pmb{{\mathscr{F}}}$ and $\mathbf{S}^\mathscr{F}=\pmb{{\mathscr{F}}}^{T} \pmb{{\mathscr{F}}}$, 
 and   $ \pmb{ \mathscr{F}} $    
 is the matrix whose columns are the $ BS $ filtered vectors  $\mathbf{ F}^b_v  $.  
 The cost of computing $\mathbf{S}^\mathscr{F} $   and  $\mathbf{H}^\mathscr{F} $  scale as  $B^2 S^2 D R^2 n $   
  and $BS( D R T n^2    +BS(DR^2 T n ))$, respectively.    
  There is no need to orthogonalize vectors or reduce their rank.  Rank reduction is by far the most time consuming step in the RRBPM of  Ref. \onlinecite{CPLec}. 
The memory cost of the calculation is determined by the memory required to store the filtered basis:  $  B S D R n $.

If 	$   \mathbf{	F}_v^b    $ were  written as a linear combination of the eigenvectors of $ \mathbf{H}  $, the dominant contributions would be from eigenvectors 
whose corresponding eigenvalues are close to $ \sigma_v^b $.  
The $ \sigma_v^b $ are distributed in a chosen energy range   $ [E_{min}$  --  $   E_{max}] $   
   and the basis of $\mathbf{	F}^b_v  $ is therefore good for the purpose of computing energy levels in the range.    
  We want the lowest energy levels, however, it  is better not to begin the range at the energy of the minimum of the PES,   but just below the ground state.   In practice, the bottom of the range is  slightly below the ground state of the harmonic Hamiltonian extracted from the full Hamiltonian by discarding all the coupling and anharmonicity.

  The MSBII eigensolver is related to Filter Diagonalization\cite{wallN95,MTdirdel97,Man98,fdzslc}, Inverse Iteration\cite{chatelin}, and Rational Krylov methods\cite{Ruhe84}.    
  We choose 
  $ \sigma_v^{b>1} $ so that their  distribution is   close to the distribution of the  eigenvalues of $ \mathbf{H}  $  in the desired range  and  choose 	 
  $\mathbf{  s}^b_v $  reasonably close to eigenvectors of   $ \mathbf{H}  $.  This has  two advantages:  1) the rank of $ \mathbf{	F}^b_v  $ is low and therefore  ALS solutions of the linear systems   are   accurate; 2)  	inner  products of 
  $	\mathbf{	F}^b_v $ with eigenvectors  of $ \mathbf{H} $ whose eigenvalues are not in the chosen energy range
  are small.    The magnitude of the    rank and the  ALS error  in 1) and the inner products in 2) 
  can  be  reduced by increasing $ P $ or $ N_{ALS} $.  However, it is always helpful  choose $\mathbf{  s}^b_v $  that are close to eigenvectors of   $ \mathbf{H}  $ and   $ \sigma_v^b $ so that their distribution approximates the distribution of the eigenvalues of $ \mathbf{H} $.
  The rank of $ \mathbf{	F}^b_v  $ is low when  it is a linear combination of a small number of exact eigenvectors, each of which is assumed to be low rank.   The method will only work if the exact eigenvectors are low rank.

    Consider first, $ b=1 $.
We choose $ S_1 $     $ \sigma_v^{b=1} $ so that they are equally spaced and within the energy range. 
We  use 
$\mathbf{  s}^{b=1}_v $  that   are nearly parallel to    low rank vectors.    
  There are $ L  $  distinct  $\mathbf{  s}^{b=1}_v $.  
 Each  distinct  $\mathbf{  s}^{b=1}_t $  $ t=1, \cdots, L $ is the sum of 
a vector   that corresponds   to  the $ t $th  energy level of the harmonic Hamiltonian referred to above 
 and  $R-1$ terms each of which is 
 an outer product of $ D $ 
 of random vectors.  
  Each component of the random vectors is a random number between 0 and 1 multiplied by $ 10^{-2}$.   It is important that the random numbers be much smaller than one. 
 This ensures that each  start vector  has significant overlap with an exact  eigenvector of $ \mathbf{H} $ and hence low rank.     
    $ S_1 $ is always chosen so that $ S_1 > L $.  
 This procedure for choosing $\mathbf{  s}^{b=1}_v $  is cheap, but yields start vectors that are linear combinations of a small number of exact eigenvectors and hence  
  low rank vectors.       
The method also works if each  start vector is a linear combination of many exact eigenvectors, but in that case it is necessary to increase $ P $ (or $ N_{ALS} $).  %
$ \sigma_v^{b=1} $ is linked to the $\mathbf{  s}^{b=1}_t $ for which the corresponding harmonic level is closest.   
Some $\mathbf{  s}^{b=1}_t $  are linked to more than one $ \sigma_v^{b=1} $ ($ S_1 > L $).    $ P=2 $ is used to obtain  the basis 
$ \pmb{\mathscr{F}}^1 $ = $ \mathbf{	F}^{b=1}_v  $, $  v=1, \cdots, S_1 $.   
Eigenvalues computed in the $ \pmb{\mathscr{F}}^1 $  basis are not accurate.   One way to improve the accuracy is to 
 choose $ P > 2 $  and increase $ S_1 $ (see section \ref{param}).   We find that it is less costly to increase $ B $ and to choose some of the     $\sigma_v^{b=2} $ equal to 
$ \pmb{\mathscr{F}}^1 $  eigenvalues.      
  If the distribution of the exact eigenvalues is non-uniform,  the equally spaced shifts used for $ b=1 $ are not ideal  for two reasons. 
   1) The filtered vector that corresponds to a shift that is far from all exact eigenvalues  will have small overlaps with all the exact eigenvectors and be a poor basis vector.  2) It is useful to have  more shifts in   dense regions of the spectrum.   
 $ P=2 $ is large enough to ensure that the distribution    of the   $ \pmb{\mathscr{F}}^1 $  eigenvalues 
 is close to the distribution of the exact eigenvalues.

For $ b=2 $,  
$ J $ of the 
  $ \sigma_v^{b=2} $  are  set equal to the eigenvalues within the range  obtained by solving the  eigenvalue problem with the the basis 
   $ \pmb{ \mathscr{F}}^1 $.    
    Additional  $ \sigma_v^{b=2} $  are chosen as follows.  
    We identify the $ S_2-J $ closest pairs of eigenvalues  obtained from the $ \pmb{ \mathscr{F}}^1  $ basis   
    and set  
     $ \sigma_{v}^{b=2} $   $ v=J+1, \cdots, S_2 $ equal to 
      the values half between these pairs.  We add      $ \sigma_{v}^{b=2} $ in the densest region(s) of the spectrum because     
  the density of shifts ought to be close to the density of the exact eigenvalues (which is  close to the density of  the  
  eigenvalues  obtained from the $ \pmb{ \mathscr{F}}^{1}  $ basis).   
  We  increase $ S_2 $ until eigenvalues of \Eq{geneig} converge.   
    The distribution of the  $ \sigma_v^{b=2} $ is  closer to the distribution of the exact eigenvalues of $ \mathbf{H} $ than is the distribution of the  $ \sigma_v^{b=1} $.   
  The $ S_2 $  $\mathbf{ s}_v^{b=2} $, start vectors are made using the same procedure as was used for  $\mathbf{ s}_v^{b=1} $,    but with different random numbers 
   and $\mathbf{	F}^{b=2}_v  $, $ v=1, \cdots, S_2 $, are computed with some chosen value of  $ P $.    
   For $ b=3, \cdots, B $,      $ \sigma_v^{b} $    = 
    $ \sigma_v^{b=2} $ and    
 $ \mathbf{s}_v^{b=2} $  are made using the same procedure as was used for  $ \mathbf{s}_v^{b=1} $, but with different random numbers, and the value of $ P $ is the same as for $ b=2. $
	We also did  calculations by updating 
$ \sigma_v^{b} $  for  $ b=3, \cdots, B $ (similar to the way we update $ \sigma_v^{b=2} $) and discovered that the results differ negligibly from those computed with 
$ \sigma_v^{b} $ = $ \sigma_v^{b=2} $ .  %

	\makeatletter
	\vspace{1cm}   
	\newenvironment{breakablealgorithm}
	{
			\refstepcounter{algorithm}
			\hrule height.8pt depth0pt \kern2pt
			\renewcommand{\caption}[2][\relax]{
				{\raggedright\textbf{\fname@algorithm~\thealgorithm} ##2\par}%
				\ifx\relax##1\relax 
				\addcontentsline{loa}{algorithm}{\protect\numberline{\thealgorithm}##2}%
				\else 
				\addcontentsline{loa}{algorithm}{\protect\numberline{\thealgorithm}##1}%
				\fi
				\kern2pt\hrule\kern2pt
			}
		}{
			\kern0pt\hrule\relax
	}
	\makeatother

\begin{breakablealgorithm}
	\caption{CP-MSBII}
	\label{MSBII}

\raggedright \textbf{Input} : $ E_{min} $   and  $ E_{max} $; shift values  $\sigma^{b=1}_v$; start vectors $\mathbf{  s}^b_v$, where $b=1, \cdots, B $ and  $ v=1, \cdots, S_1$, as 

\qquad \qquad explained in the text

\raggedright \textbf{Output} :  $BS_B$ eigenvalues and eigenvectors of    $\mathbf{  H}$   
	
	\begin{enumerate}  
		
		\item Find better shift values for $ b=2, 3, \cdots , B$
		
		\begin{enumerate}
			
			\item Loop over vectors in the  block $ b=1 $ %
			
			\textbf{for} {$v=1$ to $S_1$} 
			
			\item Loop over the number of applications of the inverse %
			
			\textbf{for} {$p=1$ to $2$}  %
			
   			    \begin{enumerate}

				 \item Call [$\mathbf{F}_v^1$] = CPII($\mathbf{  s}^1_v$,$\sigma^1_v$)

				 \item $\mathbf{  s}^1_v$ $\leftarrow$ $\mathbf{F}_v^1$
				 
				\end{enumerate}
				
		   \item Solve generalized eigenvalue problem
				
				\begin{enumerate}
					
					\item  Build matrix of filtered vectors, $\pmb{ \mathscr{F}}=[\mathbf{F}_1^1,...,\mathbf{F}_{S_1}^{1}]$    
					
					\item Compute $\mathbf{H}^{\mathscr{F}} = \pmb{ \mathscr{F}}^{T} \mathbf{H} \pmb{ \mathscr{F}}$ and $\mathbf{S}^{\mathscr{F}}=\pmb{ \mathscr{F}}^{T} \pmb{ \mathscr{F}}$.
					
					\item Solve $\mathbf{H}^{\mathscr{F}}\mathbf{U}=\mathbf{S}^{\mathscr{F}}\mathbf{U}\mathbf{E}$
					
					\item $\sigma^2_1$,..., $\sigma^2_J$  $\leftarrow$  diag$(\mathbf{E})$ (explained in the text)

					\item    Add shifts $\sigma^2_{J+1}, \cdots,    $    between the closest elements of   $ \mathbf{E}   $ (explained in the text)

				\end{enumerate}

		\end{enumerate}

		\item Other blocks
		
		\begin{enumerate}
			
			\item Loop over the number of blocks 
			
			\textbf{for} {$b=2$ to $B$}
			
			\item Loop over the vectors in the block 
			
			\textbf{for} {$v=1$ to $S_b$}
			
			\item Loop over the number of application of the inverse($P$) %
			
			\textbf{for} {$p=1$ to $P$}  
		
			\begin{enumerate}

				\item Call [$\mathbf{F}_b^v$]= CPII($\mathbf{  s}^b_v$,$\sigma^b_v$)

				\item $\mathbf{  s}^b_v$ $\leftarrow$ $\mathbf{F}_v^b$
			
			\end{enumerate}

		\end{enumerate}
		
		\item Compute eigenvalues,
		
		\begin{enumerate}
			
			\item Collect the   filtered vectors, $\pmb{ \mathscr{F}}=[\mathbf{F}_1^1,...,\mathbf{F}_{S_B}^{B}]$
			
			\item Compute $\mathbf{H}^{\mathscr{F}} = \pmb{ \mathscr{F}}^{T} \mathbf{H} \pmb{ \mathscr{F}}$ and $\mathbf{S}^{\mathscr{F}}=\pmb{ \mathscr{F}}^{T} \pmb{ \mathscr{F}}$.
			
			\item Solve $\mathbf{H}^{\mathscr{F}}\mathbf{U}=\mathbf{S}^{\mathscr{F}}\mathbf{U}\mathbf{E}$
		\end{enumerate}
		
	\end{enumerate}
	
	\hrule height.5pt depth0pt \kern2pt	 	
	Function [$\mathbf{F}$] = CPII($\mathbf{  s}$,$\sigma$) 	
	\hrule height.5pt depth0pt \kern2pt

	\begin{enumerate}  
		
		\item Initialize    
		\begin{enumerate}

			\item Assign $(\mathbf{f}_{1,\ell},..., \mathbf{f}_{D,\ell})$ $\leftarrow$ $(\mathbf{g}_{1,\ell},..., \mathbf{g}_{D,\ell})$ $\quad \forall \  \ell$

			\item Construct   $\mathbf{M}^{t}_{\forall c}$ and $\mathbf{P}_{\forall c}$ 
		\end{enumerate}
		
		\item Solve  for CP parameters 
		
		\textbf{for} {$\alpha=1 $ to $N_{ALS}$}   
		
		\textbf{for} {$j=1$ to $D$}

		\begin{enumerate}       
			\item Compute $\mathbf{M}^t_{\neq j}$ :   
			$M^{t}_{\neq j}(\ell^{\prime},\ell)=M^{t}_{\forall c}(\ell^{\prime},\ell)/M^{t}_{j}(\ell^{\prime},\ell)$
			$\quad \forall \  \ell,\ell^{\prime}$
			
			\item Compute $\mathbf{P}_{\neq j}$ :
			$P_{\neq j}(\ell^{\prime},\ell^{\prime\prime})=P_{\forall c}(\ell^{\prime},\ell^{\prime\prime})/P_j(\ell^{\prime},\ell^{\prime\prime})$ 
			$\quad \forall \  \ell^{\prime},\ell^{\prime\prime}$
			
			\item Compute $\mathbf{A}$ :           
			${A}(\ell^{\prime},i_j^{\prime},\ell,i_j)=\sum_{t=1}^{T+1}M^{t}_{\neq j}(\ell^{\prime},\ell)  h^{t}_j(i_j^{\prime},i_j)$
			$\quad \forall \ \ell^{\prime},i_j^{\prime},\ell,i_j$
			
			\item Compute $\mathbf{b}$ :
			${b}(\ell^{\prime},i_j^{\prime})=\sum_{\ell^{\prime\prime}=1}^{R}   P_{\neq j}(\ell^{\prime},\ell^{\prime\prime}) \  {g}_{i_{j}^{\prime}}^{(j,\ell^{\prime\prime})}$  %
			$\quad \forall \ \ell^{\prime},i_j^{\prime}$
			
			\item Solve the linear system for $\mathbf{x}$ :  
			${A}(\ell^{\prime},i_j^{\prime},\ell,i_j) x(\ell,i_j)={b}(\ell^{\prime},i_j^{\prime})$ 
			$\quad \forall \ \ell,i_j$
			
			\item Update $\mathbf{f}_{j,\ell}$ by replacing  
			$f_{i_j}^{(j,\ell)}    %
			\leftarrow 
			x(\ell,i_j)$ 
			$\quad \forall \ \ell,i_j$

			\item Compute $\mathbf{M}^{t}_{j}$ : 
			$M^{t}_{j}(\ell^{\prime},\ell)
			=
			{\mathbf{f}}_{j,\ell^{\prime}}^T \mathbf{h}_j^t {\mathbf{f}}_{j,\ell}$
			$\quad \forall \ \ell,\ell^{\prime},t$
			
			\item Compute $\mathbf{P}_j$ :
			$P_j(\ell^{\prime},\ell^{\prime\prime})
			=
			{\mathbf{f}}_{j,\ell^{\prime}}^T  {\mathbf{g}}_{j,\ell^{\prime\prime}}$
			$\quad \forall \ \ell^{\prime\prime},\ell^{\prime}$
			
			\item Update $\mathbf{M}^{t}_{\forall c}$ with new parameters
			$M^{t}_{\forall c}(\ell^{\prime},\ell)
			=
			M^{t}_{\neq j}(\ell^{\prime},\ell)  M^{t}_{j}(\ell^{\prime},\ell)$       
			$\quad \forall \ \ell,\ell^{\prime},t$
			
			\item Update $\mathbf{P}_{\forall c}$ with new parameters 
			$P_{\forall c}(\ell^{\prime},\ell^{\prime\prime})
			=
			P_{\neq j}(\ell^{\prime},\ell^{\prime\prime})  P_j(\ell^{\prime},\ell^{\prime\prime})$
			$\quad \forall \ \ell^{\prime\prime},\ell^{\prime}$

		\end{enumerate}  
	
	 \item Normalize $ \mathbf{F}$ $\leftarrow$  $  \frac{\mathbf{F}} {\norm {\mathbf{F}}}$

	\end{enumerate}
\end{breakablealgorithm}

 \bigskip

\vspace{1cm}

In summary, with the CP-MSBII method,
 one computes   {\emph{all}}
 states in a given energy range by solving one generalized eigenvalue problem. No vectors with rank larger than $ R $ are needed and there is  no rank reduction. 
 In contrast,  vectors with rank larger than $ R  $  and 
  rank reduction are intrinsic elements of  RRBPM calculations and the only way to use the RRBPM 
  without storing vectors whose rank is larger than $ R $ requires (see section IID of Ref. \onlinecite{CPint} and Ref.  \onlinecite{PTHS18} )   %
some additional calculations.  Although many states are calculated, there is no need to orthogonalize vectors

\section{Choosing the parameters}

\label{param}

  In practice, we choose a value of $ P $ and then select $ S_b $ and $ B $ so that energy levels are accurate.   In most calculations we have done, $ P $ is about 5.
  The basis   $ \pmb{\mathscr{F} }$ is the union of $ B $ bases each with $ S_b $ vectors.   Increasing $ B $ increases the size of the basis and therefore the accuracy of all the energies we compute. 
  Accuracy can also be improved by   making the basis vectors ``better".   This can be accomplished by increasing $ P $ or  by increasing $  N_{ALS}  $.   
  Using $ B>1 $ is particularly helpful if one desires more than about 10 states.   
  Increasing $ P $ or  $  N_{ALS}  $   reduces the   number of exact eigenvectors   that have large overlaps with a given 
  basis vector.     
 Doubling $ P $   is slightly more costly than doubling $ N_{ALS} $, but for a given filtered vector, doubling $ P $ 
 makes the 
 filtered vector ``sharper", i.e., it has significant overlap with fewer exact eigenvectors of $ \mathbf{H} $.
  We therefore use $  N_{ALS} =1 $ and $ P>1 $.  
  Increasing $ P $ decreases the value of $ B $ required to obtain accurate eigenvalues; decreasing $ P $ increases the value of $ B $ required to obtain accurate eigenvalues.  
  If $ P $ is larger, $ S_b  $  
must be larger.   If $ P $ is increased and $ S_b $ 
is not increased, it may happen that the inner products of  some exact eigenvectors of $ \mathbf{H} $ with 
$ all $  the filtered vectors are small.  In this case,
 the corresponding eigenvalues will have large errors.  If $ P $ is large  and there are no    shift values close enough to an exact eigenvalue, then it will be impossible to accurately calculate that eigenvalue.
In practice,  it is good to choose a large value of $ S_1 $.   We want to ``saturate" the energy range with shifts.  If $ S_1 $ is  too large solving the generalized eigenvalue problem might  become  difficult, but we have not encountered  problems. 
When $ P $ is too small then the filtered vectors are too ``broad", which means that a  large number  of filtered vectors 
and  a large range 
would be required to compute accurate eigenvalues.   
Increasing $ R $ never increases errors in energy levels.   However, if $P$, $S_b$, and $B$ are not chosen correctly increasing $ R $ will not yield accurate levels.

\section{Results and discussion}
\label{res}

We have tested the  CP-MSBII method by using it to compute energy levels of two Hamiltonians.  The first represents 64 
bilinearly coupled oscillators.  It has been used to test other tensor methods and is a good test because by using normal coordinates it is possible to obtain an analytic equation for the energies.\cite{CPLec,CPTree,TTVIB}  
The second is a 12-D  normal-coordinate Hamiltonian for acetonitrile.\cite{smolyak,CPLec,CPTree,TTVIB}

\subsection{64-D Bilinearly coupled model Hamiltonian}

It is not possible to store in memory a vector with  $ n^{64} $ components and therefore to use a direct product basis for a 64-D Hamiltonian, it is imperative that vectors be stored in some tensor format. 
  The Hamiltonian is
\begin{align}
\mathbf{H}
&=
\sum_{i=1}^{64} \frac{\omega_i}{2} \left( p_{i}^{2} 
+ q_{i}^{2}\right) + \sum_{i=1}^{64}\sum_{j>i}^{64} \alpha_{ij} q_{i}q_{j}
\end{align}
where $p_j=- \mathrm{i} \frac{\partial}{\partial q_{i}}$ and $\alpha_{ij}$ is a coupling constant. Following  Ref.  \onlinecite{CPLec},
we choose  $\omega_i=\sqrt{{i}/{2}}$ and  $\alpha_{ij}=0.1$. 
 There are 2144 terms in the Hamiltonian.  
The coupling constants are small, but shift energy levels significantly.

 In Table \ref{tab:64d},  the 52 lowest   energy levels   
 are reported.   
 Parameter values are in  Table \ref{64dpara}.  
 $ R=10 $
 is sufficient for computing  good approximations for all the first 52 levels.  Increasing $ R $ to 30 
 improves the accuracy of most of the energies.  Energies near the top of the chosen range are less accurate.  
 Increasing $ R $ further to 50 improves accuracy of the energies slightly.   Errors for energies near the top of the range are reduced more by increasing $ E_{max} $.
 This is  due to the fact that filtered vectors with shift values near the top of the window have significant overlaps with exact eigenvectors whose eigenvalues are outside the chosen range.   
   Energies with similar errors were computed by Rakhuba et al. \cite{TTVIB}
 using the Tensor Train  format and
 and Tensor Train rank  %
  $ R_{TT}  = 15$.   Inner products and matrix-vector products in a Tensor Train method scale as $ R_{TT}^2 $.  Their cost is similar to that of their CP counterparts if $  R_{TT}^2  = R_{CP}   $.  In our calculations,  we use  $R_{CP} = 10$ or 30.
 To do the calculation, we require less than 1 GB of memory (without a tensor format, storing a vector  with   $10^{64}$  requires $ \sim 10^{55}  $ GB ).

\begin{center}

\begin{longtable}[h]{cccccc}
	\caption{The lowest vibrational energy levels    
		of   the 64-D bilinearly coupled model Hamiltonian}  %
	\label{tab:64d}\\
	\hline
	$n$ &   %
	$E_\mathrm{exact}$ &
	$E_\mathrm{calc}(R=10)$ &
	$\frac{E_\mathrm{calc}-E_\mathrm{exact}}{E_\mathrm{exact}}$ &
	$E_\mathrm{calc}(R=30)$ &
	$\frac{E_\mathrm{calc}-E_\mathrm{exact}}{E_\mathrm{exact}}$ \\
	\endfirsthead
	\multicolumn{6}{c}
	{{\bfseries Table \thetable\ continued }} \\
	\hline
	$n$ &
	$E_\mathrm{exact}$ &
	$E_\mathrm{calc}(R=10)$ &
	$\frac{E_\mathrm{calc}-E_\mathrm{exact}}{E_\mathrm{exact}}$ &
	$E_\mathrm{calc}(R=30)$ &
	$\frac{E_\mathrm{calc}-E_\mathrm{exact}}{E_\mathrm{exact}}$ \\
	\hline
	\endhead
	\hline
0  & 121.620947675 & 121.624944552 & 3.29E-05 & 121.621022847 & 6.18E-07 \\
1  & 122.292357688 & 122.296156022 & 3.11E-05 & 122.292517450 & 1.31E-06 \\
2  & 122.585637797 & 122.600676444 & 1.23E-04 & 122.586064924 & 3.48E-06 \\
3  & 122.810589928 & 122.829500585 & 1.54E-04 & 122.811655595 & 8.68E-06 \\
4  & 122.963767701 & 122.970808029 & 5.73E-05 & 122.964708420 & 7.65E-06 \\
5  & 123.000212841 & 123.042659907 & 3.45E-04 & 123.025834375 & 2.08E-04 \\
6  & 123.167268924 & 123.187860433 & 1.67E-04 & 123.182855770 & 1.27E-04 \\
7  & 123.257047811 & 123.286986894 & 2.43E-04 & 123.267171988 & 8.21E-05 \\
8  & 123.318299395 & 123.439491265 & 9.83E-04 & 123.349524289 & 2.53E-04 \\
9  & 123.457188689 & 123.497051491 & 3.23E-04 & 123.495528490 & 3.11E-04 \\
10 & 123.481999941 & 123.527292763 & 3.67E-04 & 123.521399813 & 3.19E-04 \\
11 & 123.550327920 & 123.613924214 & 5.15E-04 & 123.571185189 & 1.69E-04 \\
12 & 123.586467177 & 123.626585991 & 3.25E-04 & 123.602569323 & 1.30E-04 \\
13 & 123.635177715 & 123.679438354 & 3.58E-04 & 123.642043976 & 5.55E-05 \\
14 & 123.671622854 & 123.714428880 & 3.46E-04 & 123.700174729 & 2.31E-04 \\
15 & 123.707892200 & 123.760056703 & 4.22E-04 & 123.741427062 & 2.71E-04 \\
16 & 123.775280050 & 123.806388473 & 2.51E-04 & 123.814687966 & 3.18E-04 \\
17 & 123.822743139 & 123.865527911 & 3.46E-04 & 123.847749617 & 2.02E-04 \\
18 & 123.838678938 & 123.892514173 & 4.35E-04 & 123.865889920 & 2.20E-04 \\
19 & 123.928457824 & 123.989597028 & 4.93E-04 & 123.971047148 & 3.44E-04 \\
20 & 123.931985750 & 123.991758896 & 4.82E-04 & 123.991696643 & 4.82E-04 \\
21 & 123.964902964 & 124.055356641 & 7.30E-04 & 124.034119090 & 5.58E-04 \\
22 & 123.989709409 & 124.041523875 & 4.18E-04 & 124.048187436 & 4.72E-04 \\
23 & 124.000232180 & 124.065808250 & 5.29E-04 & 124.049110448 & 3.94E-04 \\
24 & 124.036370297 & 124.133319041 & 7.82E-04 & 124.087447038 & 4.12E-04 \\
25 & 124.128598703 & 124.171365532 & 3.45E-04 & 124.172361270 & 3.53E-04 \\
26 & 124.131959047 & 124.184557799 & 4.24E-04 & 124.177460464 & 3.67E-04 \\
27 & 124.136493406 & 124.186369599 & 4.02E-04 & 124.195949823 & 4.79E-04 \\
28 & 124.153409954 & 124.227969314 & 6.01E-04 & 124.235559150 & 6.62E-04 \\
29 & 124.189855094 & 124.261156702 & 5.74E-04 & 124.238324082 & 3.90E-04 \\
30 & 124.221737933 & 124.295122012 & 5.91E-04 & 124.263569410 & 3.37E-04 \\
31 & 124.232838814 & 124.310919909 & 6.29E-04 & 124.310582943 & 6.26E-04 \\
32 & 124.257877190 & 124.349109296 & 7.34E-04 & 124.328252058 & 5.66E-04 \\
33 & 124.282989518 & 124.370241764 & 7.02E-04 & 124.368560923 & 6.89E-04 \\
34 & 124.306587728 & 124.384819787 & 6.29E-04 & 124.372588950 & 5.31E-04 \\
35 & 124.325805190 & 124.387249183 & 4.94E-04 & 124.376453552 & 4.07E-04 \\
36 & 124.343032868 & 124.403494874 & 4.86E-04 & 124.391574203 & 3.90E-04 \\
37 & 124.356911177 & 124.434326981 & 6.23E-04 & 124.422354037 & 5.26E-04 \\
38 & 124.379302214 & 124.438318854 & 4.74E-04 & 124.442685352 & 5.10E-04 \\
39 & 124.379478007 & 124.466253196 & 6.98E-04 & 124.466382313 & 6.99E-04 \\
40 & 124.415725719 & 124.472106328 & 4.53E-04 & 124.506299774 & 7.28E-04 \\
41 & 124.421878812 & 124.498230715 & 6.14E-04 & 124.518724305 & 7.78E-04 \\
42 & 124.446690063 & 124.544776408 & 7.88E-04 & 124.529083139 & 6.62E-04 \\
43 & 124.494153152 & 124.546455950 & 4.20E-04 & 124.582245015 & 7.08E-04 \\
44 & 124.502882243 & 124.560177758 & 4.60E-04 & 124.597988163 & 7.64E-04 \\
45 & 124.507941648 & 124.569208533 & 4.92E-04 & 124.604172798 & 7.73E-04 \\
46 & 124.510088951 & 124.647918981 & 1.11E-03 & 124.631952394 & 9.79E-04 \\
47 & 124.515018042 & 124.651712243 & 1.10E-03 & 124.645035315 & 1.04E-03 \\
48 & 124.546534091 & 124.753160978 & 1.66E-03 & 124.654108479 & 8.64E-04 \\
49 & 124.551157299 & 124.763592405 & 1.71E-03 & 124.655684714 & 8.39E-04 \\
50 & 124.587515706 & 124.832768310 & 1.97E-03 & 124.689328103 & 8.17E-04 \\
51 & 124.599867837 & 124.859193684 & 2.08E-03 & 124.722703027 & 9.86E-04

\end{longtable}

\end{center}

\begin{table}[H]
	\caption{Parameters for  the calculation  with the 64-D bilinearly coupled oscillator Hamiltonian}
	\begin{center}
		\begin{tabular}{ c c  }
			\hline
			
			Parameter      & Value \\ 
			
			\hline
			
			Rank           & 10/30 \\ 
			
			$\omega_{i}$   & $\sqrt{i/2}$ \\
			
			$\alpha_{ij}$  & 0.1  \\
			
			$n$            & 10 $\forall n_c$ \\   
			
			$N_{ALS}$	   & 1 \\	  
			
			$P$	           & 10 \\   
			
			$B$	           & 2 \\
			
			$S_1 $	       & 100  \\
			
			$S_2 $	       & 52 \\
			
			$E_{\mathrm{min}}$ & 121.60 \\
			
			$ E_{\mathrm{max}} $  & 124.60
			
		\end{tabular}				
		\label{64dpara}	    	
	\end{center}
\end{table}

\subsection{Acetonitrile(CH$_3$CN)}

It is important to   test the  CP-MSBII method on more a  realistic Hamiltonian and we have therefore computed the first 50 vibrational energy levels of  acetonitrile.  We use normal coordinates. The acetonitrile normal coordinate Hamiltonian is now      a common test problem.  Although the Hamiltonian is only 12-D, the  acetonitrile calculations are more costly.   
The Hamiltonian is 
\begin{align}
\mathbf{H}
&=
\frac{1}{2}\sum_{i=1}^{12} \omega_i \left( p_{i}^{2} +
q_{i}^{2}\right) + 
\frac{1}{6}\sum_{i=1}^{12}\sum_{j=1}^{12}\sum_{k=1}^{12} \phi^{(3)}_{ijk} q_{i}q_{j}q_{k}+
\frac{1}{24}\sum_{i=1}^{12}\sum_{j=1}^{12}\sum_{k=1}^{12}\sum_{l=1}^{12} \phi^{(4)}_{ijkl} q_{i}q_{j}q_{k}q_{l} ~.
\label{hamilt}
\end{align}
The parameters are taken from Ref. \onlinecite{smolyak} and are based on values reported in   Ref. \onlinecite{BeCaPo}, which however does not give all the parameters required for a 12-D PES.   
 The Hamiltonian operator has 323 terms:    12 kinetic energy terms, 12 quadratic potential terms, 108 cubic potential terms, and 191 quartic potential terms.
 $q_{5}$, $q_{6}$,  and $q_{7}$, $q_{8}$,  and $q_{9}$, $q_{10}$,  and $q_{11}$, $q_{12}$ are degenerate coordinates pairs.

We use the same direct product basis as Ref.  \onlinecite{CPLec}.  
The energy levels are reported in 
 Table ~\ref{tab:ch3cn}.  
 The calculation parameters are in  Table ~\ref{ch3cnpara}.
 As reference values, we use those from a pruned basis Smolyak quadrature calculation.\cite{smolyak}   %

Lower energy levels are more accurate, as was the case also for the 64-D Hamiltonian.   
Again, this is might  be  due in part to error caused by constraining the rank.     %
The filtered vectors we use as basis vectors are certainly much better than the shifted power method vectors of the RRBPM. 
 However,
 it is possible that the rank of the filtered vectors    required to compute accurate energy levels  is larger than the rank of the eigenvectors we wish to compute.
Filtered vectors for shift values near the top of the chosen range  have significant overlaps with exact eigenvectors outside the range and this will limit the accuracy of the largest eigenvalues.
Eigenvalues in a dense region of the spectrum  and degenerate eigenvalues tend to be less accurate; this is true throughout the energy range.  
It is for this reason that we put more shifts in dense regions.    
Because coupling is more important for 
 acetonitrile  than for the 64-D Hamiltonian, we use larger values of $ B, P,  $ and $ R $.

Thomas and Carrington used their H-RRBPM  to compute vibrational levels of  acetonitrile.  Their results and those of this paper are of similar accuracy
but they use a smaller rank.    It is because   H-RRBPM breaks the full problem into a sequence of lower-dimensional problems  that the H-RRBPM rank is smaller.  
An advantage of the CP-MSBII is that it never requires tensors with rank larger than $ R $.  In the H-RRBPM, tensors of large rank must be stored and then their rank is  reduced.  
 Rakhuba and Oseledets have used their TT method for   acetonitrile.   Using   $ R_{TT} = 40$, they obtain results somewhat more accurate than ours.   $ R_{TT} = 40$ corresponds to a CP rank of 1600 and we are using only $R =400$.

\begin{center}

\begin{longtable}[c]{ccrcccccc}
	\caption{Vibrational energy levels (\icm) of CH$_3$CN}  
	\label{tab:ch3cn}\\
	\hline
	Level &
	Sym. &
	\multicolumn{1}{c}{  $E_\mathrm{ref}$  }   &     
	Rank=20 &
	$E$-$E_\mathrm{ref}$ &
	Rank=200 &
	$E$-$E_\mathrm{ref}$ &
	Rank=400 &
	$E$-$E_\mathrm{ref}$ \\
	\endfirsthead
	\multicolumn{9}{c}%
	{{\bfseries Table \thetable\ continued from previous page}} \\
	\hline
	Level &
	Sym. &
	\multicolumn{1}{c}{$E_\mathrm{ref}$} & 
	Rank=20 &
	$E$-$E_\mathrm{ref}$ &
	Rank=200 &
	$E$-$E_\mathrm{ref}$ &
	Rank=400 &
	$E$-$E_\mathrm{ref}$ \\
	\hline
	\endhead
	\hline

	ZPVE                 &       & 9837.4073 & 9837.726 & 0.318 & 9837.418 & 0.011 & 9837.410 & 0.003 \\
	$\nu_{11}$           & $E$   & 360.991   & 361.08   & 0.09  & 360.99   & 0.00  & 360.99   & 0.00  \\
	&       & 360.991   & 361.09   & 0.10  & 360.99   & 0.00  & 360.99   & 0.00  \\
	$2 \nu_{11}$         & $E$   & 723.181   & 723.47   & 0.29  & 723.20   & 0.02  & 723.19   & 0.01  \\
	&       & 723.181   & 723.51   & 0.33  & 723.20   & 0.02  & 723.19   & 0.01  \\
	$2 \nu_{11}$         & $A_1$ & 723.827   & 724.14   & 0.31  & 723.85   & 0.02  & 723.84   & 0.01  \\
	& $A_1$ & 900.662   & 900.97   & 0.31  & 900.70   & 0.04  & 900.67   & 0.00  \\
	$\nu_{9}$            & $E$   & 1034.126  & 1034.39  & 0.26  & 1034.17  & 0.04  & 1034.13  & 0.01  \\
	&       & 1034.126  & 1034.49  & 0.36  & 1034.18  & 0.05  & 1034.14  & 0.01  \\
	$3 \nu_{11}$         & $A_2$ & 1086.554  & 1087.10  & 0.54  & 1086.63  & 0.08  & 1086.58  & 0.03  \\
	$3 \nu_{11}$         & $A_1$ & 1086.554  & 1087.12  & 0.56  & 1086.68  & 0.12  & 1086.58  & 0.03  \\
	$3 \nu_{11}$         & $E$   & 1087.776  & 1088.30  & 0.52  & 1087.86  & 0.08  & 1087.80  & 0.02  \\
	&       & 1087.776  & 1088.37  & 0.60  & 1088.03  & 0.25  & 1087.80  & 0.02  \\
	$\nu_{4}+\nu_{11}$   & $E$   & 1259.882  & 1260.66  & 0.78  & 1259.91  & 0.03  & 1259.89  & 0.01  \\
	&       & 1259.882  & 1260.67  & 0.78  & 1260.03  & 0.15  & 1259.90  & 0.02  \\
	$\nu_{3}$            & $A_1$ & 1388.973  & 1389.56  & 0.59  & 1389.18  & 0.21  & 1389.09  & 0.12  \\
	$\nu_{9}+\nu_{11}$   & $E$   & 1394.689  & 1395.28  & 0.59  & 1394.85  & 0.16  & 1394.75  & 0.06  \\
	&       & 1394.689  & 1395.48  & 0.79  & 1394.92  & 0.23  & 1394.83  & 0.15  \\
	$\nu_{9}+\nu_{11}$   & $A_2$ & 1394.907  & 1395.67  & 0.77  & 1395.05  & 0.15  & 1394.96  & 0.06  \\
	$\nu_{9}+\nu_{11}$   & $A_1$ & 1397.687  & 1398.34  & 0.65  & 1397.90  & 0.21  & 1397.82  & 0.13  \\
	$4 \nu_{11}$         & $E$   & 1451.101  & 1451.95  & 0.85  & 1451.23  & 0.13  & 1451.18  & 0.08  \\
	&       & 1451.101  & 1451.97  & 0.87  & 1451.25  & 0.15  & 1451.19  & 0.09  \\
	$4 \nu_{11}$         & $E$   & 1452.827  & 1453.66  & 0.83  & 1452.96  & 0.14  & 1452.89  & 0.06  \\
	&       & 1452.827  & 1453.75  & 0.92  & 1452.99  & 0.17  & 1452.90  & 0.08  \\
	$4 \nu_{11}$         & $A_1$ & 1453.403  & 1454.31  & 0.91  & 1453.58  & 0.18  & 1453.47  & 0.07  \\
	$\nu_{7}$            & $E$   & 1483.229  & 1483.73  & 0.50  & 1483.55  & 0.32  & 1483.40  & 0.17  \\
	&       & 1483.229  & 1483.76  & 0.53  & 1483.58  & 0.35  & 1483.41  & 0.18  \\
	$ \nu_{4}+2\nu_{11}$ & $E$   & 1620.222  & 1621.71  & 1.48  & 1620.64  & 0.42  & 1620.37  & 0.14  \\
	&       & 1620.222  & 1621.77  & 1.54  & 1620.73  & 0.51  & 1620.42  & 0.20  \\
	$ \nu_{4}+2\nu_{11}$ & $A_1$ & 1620.767  & 1622.55  & 1.78  & 1621.33  & 0.57  & 1620.99  & 0.22  \\
	$\nu_{3}+\nu_{11}$   & $E$   & 1749.53   & 1750.56  & 1.03  & 1749.97  & 0.44  & 1749.79  & 0.26  \\
	&       & 1749.53   & 1750.63  & 1.10  & 1750.17  & 0.64  & 1749.84  & 0.31  \\
	$ \nu_{9}+2\nu_{11}$ & $A_1$ & 1756.426  & 1757.69  & 1.26  & 1756.94  & 0.51  & 1756.69  & 0.27  \\
	$ \nu_{9}+2\nu_{11}$ & $A_2$ & 1756.426  & 1757.73  & 1.31  & 1756.97  & 0.54  & 1756.71  & 0.28  \\
	$ \nu_{9}+2\nu_{11}$ & $E$   & 1757.133  & 1758.30  & 1.17  & 1757.55  & 0.41  & 1757.32  & 0.19  \\
	&       & 1757.133  & 1758.48  & 1.35  & 1757.60  & 0.47  & 1757.34  & 0.20  \\
	$ \nu_{9}+2\nu_{11}$ & $E$   & 1759.772  & 1760.87  & 1.10  & 1760.27  & 0.50  & 1760.07  & 0.30  \\
	&       & 1759.772  & 1761.04  & 1.26  & 1760.37  & 0.60  & 1760.08  & 0.30  \\
	$2 \nu_{4}$          & $A_1$ & 1785.207  & 1787.08  & 1.88  & 1785.83  & 0.62  & 1785.45  & 0.25  \\
	$5 \nu_{11}$         & $E$   & 1816.799  & 1817.91  & 1.11  & 1817.22  & 0.42  & 1816.99  & 0.19  \\
	&       & 1816.799  & 1817.96  & 1.16  & 1817.23  & 0.43  & 1817.00  & 0.20  \\
	$5 \nu_{11}$         & $A_1$ & 1818.952  & 1820.15  & 1.19  & 1819.26  & 0.30  & 1819.12  & 0.17  \\
	$5 \nu_{11}$         & $A_2$ & 1818.952  & 1820.19  & 1.24  & 1819.28  & 0.33  & 1819.12  & 0.17  \\
	$5 \nu_{11}$         & $E$   & 1820.031  & 1821.29  & 1.26  & 1820.36  & 0.33  & 1820.16  & 0.13  \\
	&       & 1820.031  & 1821.41  & 1.37  & 1820.38  & 0.35  & 1820.19  & 0.16  \\
	$\nu_{7}+\nu_{11}$   & $A_2$ & 1844.258  & 1845.19  & 0.93  & 1844.80  & 0.54  & 1844.46  & 0.20  \\
	$\nu_{7}+\nu_{11}$   & $E$   & 1844.33   & 1845.24  & 0.91  & 1844.88  & 0.55  & 1844.59  & 0.26  \\
	&       & 1844.33   & 1845.33  & 1.00  & 1844.75  & 0.42  & 1844.61  & 0.28  \\
	$\nu_{7}+\nu_{11}$   & $A_1$ & 1844.69   & 1845.70  & 1.01  & 1845.02  & 0.33  & 1844.95  & 0.26  \\
	$\nu_{4}+\nu_{9}$    & $E$   & 1931.547  & 1933.37  & 1.83  & 1931.98  & 0.44  & 1931.76  & 0.21  \\
	&       & 1931.547  & 1933.54  & 1.99  & 1932.11  & 0.56  & 1931.78  & 0.23 
\end{longtable}

\end{center}

\begin{table}[H]
	\caption{Parameters for the  acetonitrile   calculation}
	\begin{center}
		\begin{tabular}{ c c  }
			\hline
			
			Parameter      & Values \\ 
			
			\hline
			
			Rank           & 20/200/400 \\ 
			
			$n_c$          & 9 7 9 9 9 9 7 7 9 9 27 27  \\
			
			$N_{ALS}$      & 1 \\
			
			$P$	           & 4 \\
			
			$B$	      	   & 6 \\
			
			$S_1$	       & 100 \\
			
			$S_2 $	       & 80  \\
			
			$E_{\mathrm{min}}$ & 9800 \\
				
			$ E_{\mathrm{max}} $  & 11800
			
		\end{tabular}
		\label{ch3cnpara}
	\end{center}
\end{table}

\section{Conclusion}

It is costly to solve the Schr\"odinger equation to compute a vibrational spectrum.  One problem is of course the required computer time.  Another problem is memory cost.   When the amount of memory needed to do a calculation is larger than that of the computer to which one has access, the calculation simply becomes impossible.   The simplest approach to computing a spectrum requires storing the representation of the Hamiltonian in a basis and computing some of its eigenvalues.  If $ N $
is the size of the basis then the memory cost scales as $ N^2 $.  Using iterative methods, that require only storing vectors, the memory cost scales as only $ N $. \cite{TCadv,CFSMFC12,BrWaDaCa12}   
  There are good approaches for minimizing $ N $  which  make it possible to use iterative  methods for molecules with as many as about six atoms.   They all require using a basis that is not a direct product of univariate functions.   
However, even a  memory cost that scales as  $ N $ is   debilitating.

In this paper,%
 we propose a new tensor method, the CP-MSBII.  It uses a direct product basis.   Tensor methods reduce the memory required to store wavefunctions by representing them (and all vectors used to compute them) with what is called a tensor format.   The memory cost scales linearly with $ D $.    Naturally other issues arise:  1) the algorithms become more complicated; 2) in some cases,  the computer time (not memory) required for the calculation is large.

CP-MSBII uses CP format, i.e. basis functions are simple sums of products.  Previous  tensor methods\cite{CPLec,CPTree,CPint,PTHS18,TTVIB}   
designed to simultaneously compute many states 
require orthogonalizing tensors and reducing the rank of tensors.    %
The 
CP-MSBII does not.
It uses a basis of non-orthogonal filtered vectors and solves a (small)
generalized eigenvalue problem.   
To generate the filtered vectors, we must apply $ (\sigma \mathbf{I} - \mathbf{H})^{-1} $ to vectors.   The number of applications is orders of magnitude smaller than the number of 
$ \mathbf{H} $ applications required to obtain eigenvalues of similar accuracy from the shifted block power method.   
The cost of an RRBPM calculation is mostly due to the cost of rank reduction which is not necessary when using the CP-MSBII.  
Not needing to reduce the rank also has the important advantage that there is no need to store  tensors  whose rank is larger than $ R $.   
To make standard inverse iteration work,   Rakhuba and  Oseledets \cite{TTVIB} needed to do a locally optimal
block preconditioned conjugate gradient (LOBPCG) calculation first, to prepare initial vectors.   The CP-MSBII works as is, there is no need to combine two eigensolvers.

  The accuracy of the energy levels computed with the CP-MSBII depends mostly on the choice of three parameters:  $ B, P,  N_{ALS} $.   $ P $
and $N_{ALS} $ determine the ``quality" of the basis vectors.  $ B $ determines the number of basis vectors.  We find that increasing $N_{ALS} $ is not as effective as increasing $ P $  and   set $N_{ALS} =1 $.   When $ P $ is larger, a smaller $ B $
is required and vice versa.   We must also choose the rank $ R $ and the range in which energy levels are to be determined and of course the size of the direct product basis.   We used the  CP-MSBII method to calculate  accurate vibrational energy levels of a 64-D model  Hamiltonian model and a 12-D acetonitrile Hamiltonian.   It could also be used for molecules with $ D> 12 $.

We have demonstrated that 
 the CP-MSBII method is   a good way of computing the lowest vibrational energy levels of large molecules. Often one does not want only the lowest levels or does not want to calculate {\emph{all} energy levels up to and including those of interest.
  It should be possible to use  the CP-MSBII method  also for energy levels in a window in the middle of the spectrum. 
  Regardless of where the desired levels are, it ought to be possible to speed up  CP-MSBII calculations by using a preconditioned  iterative linear solver
  to make the filtered vectors and/or by using it 
 in conjunction with a hierarchical approach.    \cite{CPTree}

\section*{Data Availability Statement}

The data that support the findings of this study are available from the corresponding author upon reasonable request.

\section*{Acknowledgements}
The financial support of the Natural Sciences and Engineering Research 
Council is gratefully acknowledged.  We thank Compute Canada for providing access to its computers.  TC thanks Mike Espig for discussions about ALS and sending him 
Ref. \onlinecite{Alscp}.  We thank Phillip Thomas and Robert Wodraszka for comments on the manuscript.

\pagebreak

\end{document}